\documentclass[a4paper,11pt]{article}
\pdfoutput=1 
\usepackage{jheppub}
\usepackage[T1]{fontenc} 
\usepackage{color}
\usepackage{fancybox}
\usepackage{amsmath}
\usepackage{amsfonts}
\usepackage{slashed}
\usepackage{amssymb}
\usepackage{indentfirst}
\usepackage{stmaryrd}
\usepackage{subcaption}
\usepackage{graphicx}
\usepackage{cleveref}
\usepackage{textcomp}
\usepackage[toc]{appendix}
\usepackage[font={small,it}]{caption}
\usepackage[justification=centering]{caption}

\allowdisplaybreaks
\newcommand{\A}{\mathbb{A}}
\newcommand{\be}{\begin{equation}}
\newcommand{\bea}{\begin{eqnarray}}

\newcommand{\ee}{\end{equation}}
\newcommand{\eea}{\end{eqnarray}}

\newcommand{\cZ}{\mathcal{Z}}

\title{\boldmath Supertwistor description of ambitwistor strings}

\author{Nathan Berkovits$^{a}$, Max Guillen$^{a,b,c}$, Lionel Mason$^{b}$}

\affiliation{$^{a}$ICTP South American Institute for Fundamental Reserch\\
Instituto de F\'{i}sica Te\'{o}rica, UNESP-Universidade Estadual Paulista\\ R. Dr. Bento T. Ferraz 271, Bl. II, S\~{a}o Paulo 01140-070, SP, Brazil\\
$^{b}$ Mathematical Institute\\
University of Oxford\\
Andrew Wiles Building, Woodstock Road, Oxford OX2 6GG, United Kingdom.\\
$^{c}$ Perimeter Institute for Theoretical Physics\\
31 Caroline St N Waterloo, Ontario N2L 2Y5, Canada
}

\emailAdd{nathan.berkovits@unesp.br,luis.max@unesp.br,lmason@maths.ox.ac.uk}

\abstract{A new ambitwistor string is constructed based on a ten-dimensional supertwistor model for the massless superparticle. Although covariant quantization is complicated by reducibility issues, a light-cone gauge analysis can be easily performed.  We show that with this analysis, this supertwistor ambitwistor string is equivalent to the  RNS ambitwistor string in light-cone gauge. In order to make the comparison, we develop the light-cone gauge analysis of the RNS ambitwistor string which has some novel features in terms of its expression of the scattering equations through interaction point operators.
}


\begin{document} 
\hfill{}
\maketitle

\section{Introduction}
Ambitwistor strings  \cite{Mason:2013sva,Berkovits:2013xba,Casali:2015vta} are chiral worldsheet theories that provide the two-dimensional quantum field theories that give rise to the CHY formulae for the scattering of massless particles in any spacetime dimension \cite{Cachazo:2013hca,Cachazo:2013iea,Cachazo:2014xea} as an extension of the four-dimensional twistor-string \cite{Witten:2003nn,Berkovits:2004hg, Skinner:2013xp}. In particular, these chiral models for Type IIA/IIB supergravity in ten-dimensional spacetime reproduce the standard tree amplitudes corresponding to the massless modes of type IIA/IIB superstring theory \cite{Adamo:2013tsa}. Yang-Mills tree amplitudes can be obtained from the chiral model of the
heterotic string, but other sectors of the heterotic ambitwistor string do not give rise to standard gravity \cite{Azevedo:2017lkz}.

In the late eighties and early nineties, there was great interest in covariant approaches to quantizing superparticle and superstring models. A novel model was introduced by one of us  \cite{BERKOVITS199045} that covariantly quantizes the ten-dimensional massless superparticle model using twistor variables. These consist of a pair of 16-component bosonic spinors of opposite chirality together with a ten-dimensional fermionic vector $(\lambda^{\alpha},w_{\beta},\psi^{m})$.  These are related to standard ten-dimensional superspace $(x^m,\theta^\alpha)$ through the \emph{incidence relations} 
\begin{eqnarray}\label{eq4}
w_{\alpha} = X_{m}(\gamma^{m}\lambda)_{\alpha} - i\psi_{m}(\gamma^{m}\theta)_{\alpha} \hspace{3mm},\hspace{3mm}\psi^{m} = (\lambda\gamma^{m}\theta)\hspace{3mm}.\hspace{3mm}
\end{eqnarray}
The twistor variables classically solve the massless condition $P^2=0$ via
\begin{equation}
P_m=\lambda^{\alpha}(\gamma_{m})_{\alpha\beta}\lambda^{\beta}\, ,\label{Plambda}
\end{equation}
where $m=0,\ldots, 9$, $\alpha,\beta=1,\ldots 16$ and $(\gamma_{m})_{\alpha\beta}$ are the 10d Pauli matrices, and the massless condition follows from the special 10d Fierz identity $(\gamma_{m})_{\alpha(\beta}(\gamma^{m})_{\gamma\delta)}=0$.  Although the solution of (\ref{Plambda})  introduces reducible constraints, covariant quantization in this twistor framework for the superparticle is
straightforward after including three generations of ghost-for-ghosts for a consistent BRST approach. Unfortunately, this twistor approach to covariantly quantizing the ten-dimensional superparticle did not generalize to the full superstring since it was unable to describe massive states.

Following the ideas developed in \cite{Mason:2013sva,Berkovits:2013xba} to construct the ambitwistor actions, we will construct the supertwistor ambitwistor string action using the ten-dimensional supertwistors mentioned above. 
 We replace the worldline of the superparticle by a Riemann sphere and the time derivatives are replaced by the antiholomorphic derivative $\bar{\partial}$ to give a  supertwistor ambitwistor string action.  This new chiral supertwistor model will be reducible  as in the superparticle version. The BRST structures of these models were studied in  \cite{Carabine:2018kdg}.
The analysis is revisited  here also in \S\ref{model} to show that the heterotic and IIB models are critical in 10d.   
The  covariant analysis is complicated with ghosts for  ghosts and so on.  To avoid these difficulties, a light-cone gauge analysis will be performed here. 

We first develop the RNS model of \cite{Mason:2013sva} in light-cone gauge to obtain new formulae for tree amplitudes.  These are still based on solutions to the scattering equations, but these are presented in a novel form in terms of interaction operators.  We then give a light-cone gauge treatment of the twistorial 10d ambitwistor-string.  In this gauge we are able to define  physical vertex operators and interaction-point operators. Using the standard light-cone gauge amplitude prescription, we demonstrate that this formalism is equivalent to the light-cone RNS ambitwistor string framework and can be used to compute tree amplitudes. Work is in progress on providing a fully covariant description of this ten-dimensional supertwistor model.

This paper is organized as follows. In section 2 we review the supertwistor description of the ten-dimensional massless superparticle starting from the ten-dimensional Brink-Schwarz superparticle, and ten-dimensional super-Maxwell is shown to be described by canonical first quantization of this model. In section 3 we introduce the supertwistor ambitwistor string using the action found in section 2. We then give a light-cone gauge treatment of the RNS ambitwistor string in section 4.  Using a similar light-cone gauge choice, we fix in section 5 all the  constraints in the supertwistor ambitwistor string and demonstrate the equivalence of N-point tree-level scattering amplitudes in this formalism with N-point tree-level scattering amplitudes in the light-cone gauge RNS ambitwistor string.

\section{Review of supertwistors for $10D$ massless superparticles}
\subsection{Standard $10D$ massless superparticle}
The ten-dimensional Brink-Schwarz superparticle is described by the action:
\begin{eqnarray}\label{eq1}
S &=& \int d\tau [P_{m}(\dot{X}^{m} - i\dot{\theta}\gamma^{m}\theta) + \frac{1}{2}e P^{2}]
\end{eqnarray}
where $X^{m}$, $P^{m}$ are bosonic $SO(1,9)$ vectors, $\theta^{\alpha}$ is a fermionic $SO(1,9)$ Majorana-Weyl spinor, $e$ is the Lagrange multiplier enforcing the massless condition $P^2=0$ and $(\gamma^{m})_{\alpha\beta}$, $(\gamma^{m})^{\alpha\beta}$ are the Pauli matrices, symmetric real 16$\times$ 16 matrices satisfying $$(\gamma^{m})
^{\alpha\beta}(\gamma^{n})_{\beta\delta} + (\gamma^{n})^{\alpha\beta}(\gamma^{m})_{\beta\delta} = 2\eta^{mn}\delta^{\alpha}_{\delta}.$$

This action is invariant under the global Poincar\'e group 
together with the global supersymmetry:
\begin{eqnarray}
\delta \theta^{\alpha} = \epsilon^{\alpha} \,,\hspace{4mm} \delta X^{m} = -i(\delta\theta\gamma^{m}\theta) \, ,\hspace{4mm} \delta P^{m} = 0 \,,\hspace{4mm} \delta e = 0\, .
\end{eqnarray}
with conserved currents for the super-Poincar\'e group, $p_m:=P_m$ for translations and
\begin{eqnarray}
M^{mn} &:=& \frac{1}{2}P^{[m}X^{n]} + \frac{i}{4}P_{p}(\theta\gamma^{mnp}\theta)\, , 
\\
 q_{\alpha} &:=& -2i P_{m}(\gamma^{m}\theta)_{\alpha}\, ,
\end{eqnarray}
for Lorentz transformations and supersymmetry respectively.

The action also has a local fermionic \emph{$\kappa$-symmetry}:
\begin{eqnarray}
\delta\theta^{\alpha} = P^{m}(\gamma_{m}\kappa) \hspace{4mm},\hspace{4mm} \delta X^{m} = -i(\theta\gamma^{m}\delta\theta) \hspace{4mm},\hspace{4mm} \delta e = -4i \dot{\theta}^{\alpha}\kappa_{\alpha} \hspace{4mm},\hspace{4mm} \delta P^{m} = 0\, ,\label{kappa}
\end{eqnarray}
and a gauge symmetry
\begin{equation}
\delta e=\dot \epsilon\, , \quad \delta X^m=-\epsilon P^m, \quad \delta (\theta^\alpha,P_m)=0.\label{P2}
\end{equation}
The orbits of these two symmetries together are super null geodesics of dimension $1|8$ and reducing $(X^m,P_m,\theta^\alpha)|_{P^2=0}$ by these local symmetries gives Witten's superambitwistor space $\A$, the $18|8$-dimensional phase space of massless 10d-superparticles \cite{Witten:1985nt}.

\subsection{Review of supertwistor description of the $D=10$ massless superparticle}
We define a supertwistor to be  $\cZ=(\lambda^\alpha,w_\alpha,\psi_m)$ where $\psi^{m}$ is a fermionic real ten-dimensional vector and the bosonic parts
$\lambda^{\alpha}$ and $w_{\beta}$ are real 16 component spinors of opposite chirality combining to form a bosonic twistor $Z_A$, a 32 component chiral spinor for the conformal group $SO(2,10)$.  There is a natural invariant skew form on such supertwistors
\begin{equation}
\Omega(\cZ_1,\cZ_2)=\lambda_1^\alpha w_{2\alpha}-\lambda_2^\beta w_{1\beta}+i\psi_1^m\psi_{2m}-i\psi_2^m\psi_{1m}\, .
\end{equation}

In order to describe the ten-dimensional superparticle using  supertwistors,
one solves the massless condition $P^{2} = 0$ using \eqref{Plambda} to define $P_m$ in terms of $\lambda^\alpha$, $ P^{m} = (\lambda\gamma^{m}\lambda)$. It then follows  that  
\begin{eqnarray}\label{eq4}
w_{\alpha} = X_{m}(\gamma^{m}\lambda)_{\alpha} - i\psi_{m}(\gamma^{m}\theta)_{\alpha} \hspace{3mm},\hspace{3mm}\psi^{m} = (\lambda\gamma^{m}\theta)\hspace{3mm},\hspace{3mm}
\end{eqnarray}
are invariant under the $\kappa$-symmetry \eqref{kappa} and gauge symmetry \eqref{P2}.  In order to be able to obtain $(X^m,\theta^\alpha)$ satisfying \eqref{eq4}, $\cZ$ must be  subject to the constraints
\begin{eqnarray}
g &:=& (\lambda\gamma^{m}\lambda)\psi_{m} = 0\\
G^{\alpha} &:=& (\lambda\gamma^{m}\lambda)(\gamma_{m}w)^{\alpha} - 2\lambda^{\alpha}(\lambda w) + 2i\psi^{m}\psi^{n}(\gamma_{m}\gamma_{n}\lambda)^{\alpha} = 0\,.
\end{eqnarray}
These constraints are not independent of each other as
\begin{eqnarray}
H^{m} := (\lambda\gamma^{m}G) - 4i\psi^{m}g = 0
\end{eqnarray}
Using the $10d$-gamma matrix identity $(\gamma_{m})_{(\alpha\beta}(\gamma^{m})_{\delta)\epsilon} = 0$, one readily finds that $(\lambda\gamma^{m}\lambda)H_{m} = 0$. Thus one is left with 16 - 9 = 7 independent bosonic constraints. These first-class constraints generate the gauge transformations 
\begin{eqnarray}
\delta_{\eta} w_{\alpha} &=& 2(\gamma^{m}\lambda)_{\alpha}(\eta\gamma_{m}w) - 2\eta_{\alpha}(\lambda w) - 2(\lambda\eta)w_{\alpha} + 2i \psi^{m}\psi^{n}(\eta\gamma_{m}\gamma_{n})_{\alpha}\nonumber\\
\delta_{\eta}\lambda^{\alpha} &=& - (\gamma^{m}\eta)^{\alpha}(\lambda\gamma_{m}\lambda) + 2(\lambda\eta)\lambda^{\alpha}\nonumber\\
\delta_{\eta}\psi^{m} &=& \psi^{n}(\eta\gamma_{n}\gamma^{m}\lambda) - \psi^{n}(\eta\gamma^{m}\gamma_{n}\lambda)\nonumber \\
\delta_{\xi}w_\alpha  &=& 2 \xi(\gamma^{m}\lambda)_{\alpha} \psi^m \nonumber\\
\delta_{\xi}\psi^m &=& \xi(\lambda\gamma^m \lambda)\label{eq46}
\end{eqnarray}
where $\eta_{\alpha}$ and $\xi$ are arbitrary $SO(1,9)$ bosonic spinor and fermionic scalar parameters respectively. So the twistor model actually possesses $32 - 14 = 18$ independent bosonic and $10 - 2 = 8$ independent fermionic degrees of freedom, i.e., the dimension of $\A$, the phase space of the ten-dimensional Brink-Schwarz superparticle.

The above relations imply
\begin{eqnarray}
\dot{X}^{m}P_{m}&=& 2\lambda^{\alpha}\dot{w}_{\alpha} + 2i\dot{\psi}^{m}\psi_{m} + 2i\psi^{m}(\lambda\gamma_{m}\dot{\theta}) - \partial_{\tau}(X^{m}P_{m})\nonumber\\
-iP^{m}(\dot{\theta}\gamma_{m}\theta) &=& -2i\psi^{m}(\lambda\gamma_{m}\dot{\theta})
\end{eqnarray}
So ignoring boundary terms, the superparticle action \eqref{eq1} can be written in terms of supertwistor variables as
\begin{equation}
\label{eq5}
S = \int d\tau [\Omega(\cZ,\dot\cZ) + h_{\alpha}G^\alpha +fg]= \int d\tau[2\lambda^{\alpha}\dot{w}_{\alpha} + 2i\dot{\psi}^{m}\psi_{m} + h_{\alpha}G^\alpha
 + f g]
\end{equation}
where $h_{\alpha}$, $f$ are Lagrange multipliers enforcing the twistor constraints. 

The super-Poincare currents can be written in terms of supertwistors as 
\begin{eqnarray}
p_{m} = (\lambda\gamma_{m}\lambda)\hspace{2mm},\hspace{2mm} q_{\alpha} = 4i \psi^{m}(\gamma_{m}\lambda)_{\alpha}\hspace{2mm},\hspace{2mm} M^{mn} &=& \frac{1}{2}(\lambda\gamma^{mn}w) - \frac{i}{2}\psi^{[m}\psi^{n]}
\end{eqnarray}
where  the Lorentz generators are obtained from the identity 
\begin{eqnarray}
(\lambda\gamma^{n}\gamma^{p}w) &=& \eta^{np}X_{m}P^{m} + 2X^{[p}P^{n]} - i\psi_{m}(\lambda\gamma^{n}\gamma^{p}\gamma^{m}\theta)
\end{eqnarray}
that follows from \eqref{eq4}.

\subsection{Quantization}
The canonical quantization yields the (anti)commutators for the superwistor variables
\begin{eqnarray}
[\lambda^{\alpha} , w_{\beta}] = \frac{i}{2}\delta^{\alpha}_{\beta} \hspace{4mm},\hspace{4mm} \{\psi^{m} , \psi^{n}\} = -\frac{1}{4}\eta^{mn}
\end{eqnarray}
Therefore the $\psi^{m}$ operators will be represented by $SO(1,9)$ $\Gamma$-matrices and the superparticle wavefunction will be described by an $SO(1,9)$ 32-component spinor $\phi_A$. The supertwistor constraints in a $\phi_A(\lambda)$ representation take the form
\begin{eqnarray}
G_{A }{}^{\alpha B}\phi_{B} &:=& \frac{1}{2i}[(\lambda\gamma^{m}\lambda)(\gamma_{m})^{\alpha\beta}\frac{\partial}{\partial\lambda^{\beta}} - 2\lambda^{\alpha}(\lambda^{\beta}\frac{\partial}{\partial\lambda^{\beta}})]\phi_{A} - \nonumber \\
&&\qquad\qquad\frac{i}{4}(\gamma^{m}\gamma^{n}\lambda)^{\alpha}(\Gamma_{m}\Gamma_{n})_{A}{}^{C}\phi_{C} - 2i\lambda^{\alpha}\phi_{A} = 0
\label{eq6}\\
g_{A}{}^{B}\phi_{B} &:=& \frac{i}{2\sqrt{2}}(\lambda\gamma^{m}\lambda)(\Gamma_{m})_{A}{}^{B}\phi_{B} = 0\label{eq77}
\end{eqnarray}
where $(\Gamma_{m})_{A}{}^{B}$ is an $SO(1,9)$ $32 \times 32$ gamma matrix. The last term in \eqref{eq6} comes from normal ordering ambiguities  and is fixed by requiring $(\lambda\gamma^{m})_{\beta}G_{A}{}^{\beta\,B} + i\sqrt{2}(\Gamma^{m})_{A}{}^{C}g_{C}{}^{B} = 0$. 

These constraints can be solved using the chiral components of the spinor 32-component $\phi_{A}=(\phi_{\alpha},\phi^{\beta})$ as 
\begin{eqnarray}
\phi_{\alpha} &=& A_{m}(\gamma^{m}\lambda)_{\alpha}\label{eq8}\\
\phi^{\beta} &=& -2\sqrt{2}(B_{\alpha}\lambda^{\alpha})\lambda^{\beta} + \frac{\sqrt{2}}{2}(B\gamma^{m})^{\beta}(\lambda\gamma_{m}\lambda)\label{eq9}
\end{eqnarray}
where $A_{m}(\lambda\gamma^{m}\lambda) =0$ and $A_m$ and $B_\alpha$ are functions only of the momentum $(\lambda\gamma^m\lambda)$. Note that $\phi_{\alpha}$ is invariant under the transformation $A_{m}\rightarrow A_{m} + C(\lambda\gamma_{m}\lambda)$, and $\phi^{\alpha}$ is invariant under the transformation $B_{\alpha} \rightarrow B_{\alpha} + (\lambda\gamma^{m}\lambda)(\gamma_{m})_{\alpha\beta}F^{\beta}$, for arbitrary $C$ and $F^{\beta}$. The gauge invariant object constructed out of $B_{\alpha}$ given by
\begin{eqnarray}
C^{\alpha} &=& (\lambda\gamma^{m}\lambda)(\gamma_{m}B)^{\alpha}
\end{eqnarray}
satisfies the usual Dirac equation in momentum space $(\lambda\gamma^{m}\lambda)(\gamma_{m}C)_{\alpha} = 0$.

A supersymmetric vertex can be obtained from \eqref{eq8}, \eqref{eq9} using the 32 component spinor ground state. In the Weyl representation, this state can be divided into two 16 component spinors $\vert 0\rangle ^\alpha$ and $\vert 0\rangle _\alpha$ where the supersymmetry generators written in matrix notation are
\begin{eqnarray}
(q_{\alpha})^{\delta\epsilon} &=& -\sqrt{2}(\gamma^{m})^{\delta\epsilon}(\gamma_{m}\lambda)_{\alpha}\nonumber\\
(q_{\alpha})_{\delta\epsilon} &=& -\sqrt{2}(\gamma^{m})_{\delta\epsilon}(\gamma_{m}\lambda)_{\alpha}
\end{eqnarray} 
Using the commutation relations 
\begin{eqnarray}
[q_{\alpha}, A_{m}] = (\gamma_{m})_{\alpha\beta}C^{\beta} \hspace{2mm}&,&\hspace{2mm} [q_{\alpha},B_{\beta}] = (\gamma^{m})_{\alpha\beta}A_{m}
\end{eqnarray}
and the action of the operator $q_{\alpha}$ on the ground state
\begin{eqnarray}
(q_{\alpha}\vert 0\rangle )^{\beta} = -\sqrt{2}(\gamma^{m}\lambda)_{\alpha}(\gamma_{m})^{\beta\delta}\vert 0\rangle _{\delta}\hspace{2mm}&,&\hspace{2mm} (q_{\alpha}\vert 0\rangle )_{\beta} = -\sqrt{2}(\gamma^{m}\lambda)_{\alpha}(\gamma_{m})_{\beta\delta}\vert 0\rangle ^{\delta}
\end{eqnarray}
one finds that the state $V$ defined to be
\begin{eqnarray}
V &=& -(\gamma^{m}\lambda)_{\beta}A_{m}\vert 0\rangle ^{\beta} - 2\sqrt{2}(B_{\beta}\lambda^{\beta})\lambda^{\delta}\vert 0\rangle _{\delta} + \frac{\sqrt{2}}{2}(\lambda\gamma_{p}\lambda)(B\gamma^{p})^{\delta}\vert 0\rangle _{\delta}
\end{eqnarray}
is supersymmetric invariant.

\section{Supertwistor $10D$ ambitwistor strings}\label{model}
The supertwistor ambitwistor action for the heterotic superstring, based on the superparticle action \eqref{eq5}, 
 will be defined on the Riemann surface $\Sigma$ to be
\begin{eqnarray}
S_{Het.} =\int_\Sigma \left(\Omega(\cZ,\bar{\partial}\cZ)
+ h_{\alpha}G^\alpha
+ f g
+ b\bar{\partial} c\right) + S_{J}\label{eq80}
\end{eqnarray}
where $S_{J}$ stands for a current algebra action as in the standard $SO(32)$ or $E_8\times E_8$ heterotic superstring.  Here now $\cZ$ are taken to be spinors in $K^{1/2}$, where $K$ is the bundle of holomorphic 1-forms on the worldsheet. The Lagrange multiplier/gauge fields
$h_\alpha$ and $f$ are  $(0,1)$ forms with values in  $K^{-3/2}$. 

 Similarly, the ambitwistor action for the Type IIB superstring is defined by doubling up the fermionic coordinates to obtain 
\begin{multline}
S_{IIB} = \int d^{2}z \left(w_{\alpha}\bar{\partial}\lambda^{\alpha} + \frac{1}{2}\psi^{m}\bar{\partial}\psi_{m} + \frac{1}{2}\tilde{\psi}^{m}\bar{\partial}\tilde{\psi}_{m} + f(\lambda\gamma^{m}\lambda)\psi_{m} + \tilde{f}(\lambda\gamma^{m}\lambda)\tilde{\psi}_{m}\right.\\
\left. + h_{\alpha}[(\lambda\gamma^{m}\lambda)(\gamma_{m}w)^{\alpha} -2\lambda^{\alpha}(\lambda w) + \psi^{m}\psi^{n}(\gamma_{m}\gamma_{n}\lambda)^{\alpha} + \tilde{\psi}^{m}\tilde{\psi}^{n}(\gamma_{m}\gamma_{n}\lambda)^{\alpha}] + b\bar{\partial} c\right)
\end{multline}
where $\tilde \psi_m$ is a second fermionic vector and the incidence relation \eqref{eq4} becomes 
\begin{eqnarray}
w_{\alpha} &=& (\gamma_{m}\lambda)_{\alpha}X^{m} + (\gamma_{m}\theta)_{\alpha}\psi^{m} + (\gamma_{m}\tilde{\theta})_{\alpha}\tilde{\psi}^{m}
\end{eqnarray}
The $N=2$ $D=10$ supersymmetry generators are $q_\alpha = \int dz \psi^m (\gamma_m\lambda)_\alpha$ and $\tilde q_\alpha = \int dz
\tilde\psi^m (\gamma_m \lambda)_\alpha$ which have the same spacetime chirality, so the superstring is type IIB and there surprisingly does not seem to be a
type IIA version of this ambitwistor action. To simplify notation, we will focus in the rest of this paper on the heterotic model given in eqn. \eqref{eq80}, however,
all results are expected to easily generalize to the Type IIB model.

The OPEs satisfied by the canonical variables are given by
\begin{eqnarray}\label{eq7}
\lambda^{\alpha}(z)w_{\beta}(w) &\rightarrow & \frac{\delta^{\alpha}_{\beta}}{z-w}\\
\psi^{m}(z)\psi^{n}(w) &\rightarrow & \frac{\eta^{mn}}{z-w}
\end{eqnarray}
and the energy-momentum tensor is
\begin{eqnarray}
T_{B}(z) &=& \frac{1}{2}\partial w_{\alpha}\lambda^{\alpha} - \frac{1}{2}w_{\alpha}\partial\lambda^{\alpha} - \frac{1}{2}\psi^{m}\partial\psi_{m} + T_{J}
\end{eqnarray}
where $T_{J}$ is the stress-energy tensor associated to the current algebra.  Using $T_{B}$ defined above,  the central charges corresponding to the $S_{\psi\psi}$ and $S_{\lambda w}$ systems are
\begin{eqnarray}
c_{\psi\psi} = \frac{D}{2}\hspace{2mm},\hspace{2mm} c_{\lambda w} = 4-2D
\end{eqnarray}

Denote the scalar and spinor constraints $T_{F}$ and $G^{\alpha}$ respectively
\begin{eqnarray}
T_{F} &=& (\lambda\gamma^{m}\lambda)\psi_{m}\\
G^{\alpha} &=& (\lambda\gamma^{m}\lambda)(\gamma_{m}w)^{\alpha} -2\lambda^{\alpha}(\lambda w) + \psi^{m}\psi^{n}(\gamma_{m}\gamma_{n}\lambda)^{\alpha}
\end{eqnarray}

Using eqns. \eqref{eq7}, one finds the constraint algebra to be
\begin{eqnarray}
G^{\alpha}(z)T_{F}(w) &\rightarrow & -\frac{2}{(z-w)}\lambda^{\alpha}T_{F}(w)\nonumber\\
T_{B}(z)G^{\alpha}(w) &\rightarrow & \frac{3}{2(z-w)^{2}}G^{\alpha}(w) + \frac{1}{(z-w)}\partial G^{\alpha}(w)\nonumber\\
T_{B}(z)T_{F}(w) &\rightarrow & \frac{3}{2(z-w)^{2}}T_{F}(w) + \frac{1}{(z-w)}\partial T_{F}(w)\nonumber\\
T_{B}(z)T_{B}(w) &\rightarrow & \frac{-\frac{1}{2}(2D-4) + \frac{D}{4} + \frac{c_{J}}{2}}{(z-w)^{4}} + \frac{2}{(z-w)^{2}}T_{B}(w) + \frac{1}{(z-w)}\partial T_{B}(w)\nonumber\\
T_{F}(z)T_{F}(w) &\rightarrow & regular\nonumber\\
G^{\alpha}(z)G^{\beta}(w) &\rightarrow & -\frac{4}{(z-w)}\lambda^{[\alpha}G^{\beta]} - \frac{56}{(z-w)^{2}}\lambda^{\alpha}\lambda^{\beta} - \frac{36}{(z-w)}\partial\lambda^{\beta}\lambda^{\alpha}\nonumber\\
&& -\frac{20}{(z-w)}\partial\lambda^{\alpha}\lambda^{\beta} + \frac{16}{(z-w)^{2}}(\gamma^{m})^{\alpha\beta}(\lambda\gamma_{m}\lambda) + \frac{16}{(z-w)}(\partial\lambda\gamma^{m}\lambda)(\gamma_{m})^{\alpha\beta}\nonumber\\\label{eq10}
\end{eqnarray}
In principle one might use \eqref{eq10} to construct the BRST operator and the corresponding BRST-closed vertex operators. However this task is not so simple, since the supertwistor ambitwistor string is a reducible constrained system where $G^{\alpha}$ and $T_{F}$ are related to each other through the relation
\begin{eqnarray}
(\lambda\gamma^{m}G) - 2\psi^{m}T_{F} &=& 0
\end{eqnarray}
and the coefficients of this relation are in turn constrained to obey
\begin{eqnarray}
(\lambda\gamma^{m}\lambda)(\gamma_{m}\lambda)_{\alpha} = (\lambda\gamma^{m}\lambda)\psi_{m} = 0
\end{eqnarray}
This implies three generations of ghosts, which will give rise to heavy algebraic manipulations. For instance, the BRST operator up to the first ghost generation is
\begin{eqnarray}
Q &=& \int dz [c T_{B} + \gamma T_{F} + c_{\alpha}G^{\alpha} + b c\partial c + \frac{3}{4}\partial c \gamma\beta + \frac{1}{4}c \gamma\partial\beta - \frac{3}{4}c\partial\gamma \beta - 2 c_{\alpha}\lambda^{\alpha}\gamma\beta - \frac{3}{4}\partial c c_{\alpha}b^{\alpha}\nonumber\\
&& - \frac{1}{4}c c_{\alpha}\partial b^{\alpha} + \frac{3}{4}c\partial c_{\alpha}b^{\alpha}+ 2\lambda^{\alpha}c_{\alpha}c_{\beta}b^{\beta} + [((\lambda\gamma^{m}b) + 2\beta\psi^{m})\gamma_{m} + (\lambda\gamma^{m}\lambda)\tilde{c}\beta_{m}] + \ldots]\nonumber\\
\end{eqnarray}
where $\ldots$ stands for contributions coming from the next ghost generations. The ghost pairs of these generations have been denoted by ($c$, $b$), ($\gamma$, $\beta$), ($c_{\alpha}$, $b^{\alpha}$), ($\gamma_{m}$, $\beta^{m}$), ($\tilde{c}$, $\tilde{b}$) and one can easily calculate the total matter and ghost central charge to be\footnote{The $2^{\frac{D}{2}-1}$ terms arise as the dimensions of the chiral spin spaces in general dimension. }
\begin{eqnarray}\label{eq47}
c_{Total} = -2^{\frac{D}{2}-1} + \frac{D}{2} - 26 + 11 - 11 \times 2^{\frac{D}{2}-1} + 26D - 74 + c_{J} 
%
\nonumber
\end{eqnarray}
So when $D=10$, cancellation of the conformal anomaly implies $c_J=16$ as in the $E_{8}\times E_{8}$ or $SO(32)$ heterotic models. The type IIB model can also be readily shown to be free of conformal anomalies in $D=10$ since
\begin{eqnarray}
c_{Total} = -2^{\frac{D}{2}-1} + \frac{D}{2}+ \frac{D}{2} - 26 + 11+11 - 11 \times 2^{\frac{D}{2}-1} + 26D - 74 
\end{eqnarray}
which again vanishes in D=10 \cite{Carabine:2018kdg}.

To avoid the algebraic complications arising from covariant quantization, we will perform a light-cone gauge analysis here which will require the gauge-fixing of the symmetries generated by $G^{\alpha}$ and the stress-energy tensor. The covariant quantization of the supertwistor ambitwistor string will hopefully be addressed in a forthcoming paper.

\section{Light-cone gauge RNS ambitwistor string}\label{apA}
The light-cone gauge description of the RNS superstring was introduced in the early stages of the construction of string theory and was mainly developed by Mandelstam in \cite{Mandelstam:1985ww,MANDELSTAM1973205}. On the other hand, the RNS ambitwistor string was recently constructed in \cite{Mason:2013sva}, where it was interpreted as the infinite tension limit of the standard RNS string. Following the same line of reasoning used in constructing the  RNS string in light-cone gauge, we  formulate a light-cone gauge quantization of the RNS ambitwistor string.   The original heterotic RNS ambitwistor string, ignoring the current algebra variables in $S_J$, has action
\begin{equation}
S_{RNS}=\int_\Sigma P_m\bar\partial X^m + \frac12 \Psi_m\bar \partial\Psi^m- \frac12 e P_mP^m -\tilde{e}(P\cdot \partial X+\Psi\cdot\partial\Psi) -\chi P_m\Psi^m\, ,
\end{equation}
where $e$, $\tilde e$ and $\chi$ are the Lagrange multipliers for the constraints $$
P_mP^m=0,\quad P\cdot \partial X+\Psi\cdot\partial\Psi=0,\quad
P_m\Psi^m=0
$$ respectively and we will take the Riemann surface $\Sigma$ to be the Riemann sphere $\mathbb{CP}^1$.  These Lagrange multipliers are also gauge fields generating symmetries
\begin{multline}\label{para}
\delta(X^m,P_m,\Psi^m,e, \tilde e,\chi)=\\(\alpha P^m+\epsilon \Psi^m +\tilde\alpha\partial X^m, \partial (\tilde \alpha P_m),\epsilon P^m +\sqrt{\tilde\alpha}\partial(\sqrt{\tilde\alpha}\Psi^m),\bar\partial \alpha ,\bar\partial\tilde{\alpha},\bar\partial \epsilon)
\end{multline}
where $\alpha$, $\tilde\alpha$ and $\epsilon$ are respectively two bosonic and one fermionic gauge symmetry parameters, $\tilde \alpha$ corresponding to infinitesimal holomorphic coordinate transformations.

We will quantize in light-cone gauge where all ghosts and non-physical variables decouple. 
Decompose 10-vectors to $1+1+8$, with $i=1,\ldots,8$ so that
\begin{equation}\label{dec}
P^m=(P^+,P^-,P^i)\, ,\qquad P_mP^m=-2P^+P^- + P^iP^i\, ,
\end{equation}
We first use the symmetries parameterized by $\alpha$ and $\epsilon$ in (\ref{para})
of the RNS ambitwistor-string  to gauge $X^+=\Psi^+=0$.
Through the equation of motion for $P^-$, this implies that
$e P^+ =0$. We will assume that $P^+$ is nonzero, so that this gauge implies $e=0$.
Similarly, the equation of motion for $\Psi^-$ implies that $\chi=0$.
The equations of motion for $e$ and $\chi$ in this gauge imply that 
\begin{equation}
P^-=\frac{P^iP^i}{2P^+}\, , \qquad \Psi^-=\frac{P_i\Psi^i}{P^+}
\, . \label{LC-solve}
\end{equation}
Although these equations potentially introduce poles into $P^-$ and $\Psi^-$ at the zeroes of $P^+$,  we will later find that the interaction point operators inserted at the zeroes of $P^+$ involve delta functions that set the residues of the poles to zero.

The action is now reduced to
\begin{equation}
S_{RNS}=\int [-P^+\bar\partial X^-+ P^i\bar\partial X^i + \frac12 \Psi^i\bar \partial\Psi^i-\tilde{e}(-P^+\partial X^- + P^i \partial X^i+\Psi^i\cdot\partial\Psi^i)] \, .
\end{equation}
There are two gauge-fixing choices of the remaining symmetries parametrized by $\tilde\alpha$ in (\ref{para}) that we will wish to bear in mind. The first follows by setting $\tilde e=0$ so that the corresponding  coordinate $z$ is a standard affine coordinate on the Riemann sphere. In the presence of vertex operators with exponential factors $e^{k_r\cdot X}$, these exponentials can be taken into the action to provide sources for $(P^+,P_i)$ giving the equations of motion for $(P^+(z),P_i(z))$  
\begin{equation}
\bar\partial P^+=\sum_r  k_r^+\,\delta^2(z-z_r)\,
\quad
\bar\partial P^i=\sum_r  k^{i}_{r}\,\delta^2(z-z_r)\, .
\end{equation}
These have the unique solutions
\begin{eqnarray}
P^+(z) = \sum_{r=1}^{N}\frac{k^+_r}{z-z_{r}}\, ,
\quad
P^{i}(z)= \sum_{r=1}^{N}\frac{k_{r}^{i}}{z-z_{r}}\, ,
\end{eqnarray}
where $k^m_r = (k^+_r, k^-_r, k^i_r)$ are the momenta of the external states.

The standard light cone coordinate $\rho=\sigma+i\tau$ for the conventional string identifies $X^+=\tau$ and this has the effect of setting $P^+_\rho=1$. In the ambitwistor string, the string lies in the space of complex null geodesics, ambitwistor space $\mathbb{A}$, and in light cone gauge we are choosing coordinates $(P^+,P^i, X^-,X^i)$  on $\mathbb{A}$  where $(X^-,X^i)$ is the point where the geodesic intersects $X^+=0$ and $(P^+,P^i)$ parametrizes its null momentum.  Although we cannot therefore identify $\tau$ with $X^+$, we can nevertheless make an alternative choice of
the gauge-fixing for $\tilde\alpha$ by imposing the condition 
\begin{equation}
P_\rho^+=1\, .
\end{equation}
With this gauge choice, the equation of motion for $X^-$ implies $\partial \tilde e=0$
and the constant mode of $\tilde e$ acts as a Lagrange multiplier for the remaining
light-cone constraint $\int dz (P^j \partial X^j + \Psi^j \partial \Psi^j)$, which is the
usual $L_0 - \bar L_0$ condition.

When expressed in terms of the $z$ coordinate defined using the first gauge-fixing choice,
\begin{equation} 
P^m_\rho d\rho=P^m_zdz\, .
\end{equation} This implies 
 $$
\frac{\partial \rho}{\partial z}=P^+_\rho\frac{\partial\rho}{\partial z}=P^+_z(z)=\sum_r \frac{k^+_r}{z-z_r}\, ,
$$
so one arrives at the usual Mandelstam map 
\begin{equation}
\rho=\sum_r k^+_r \log (z-z_r)
\end{equation}
relating the light-cone coordinate with the Riemann sphere.
In this coordinate, strings come in from infinity as cylinders in the $\rho$ coordinate corresponding to the points $z=z_r$ in conformal gauge and join in pairs of pants at the $n-2$ interaction points $\tilde z_\alpha$  where $P^+_z=0$.   In light-cone gauge, the differences of these interaction points, $\rho(\tilde z_\alpha)-\rho(\tilde z_1)$  for $\alpha =2$ to $n-2$, naturally parametrise the $n-3$ moduli of the $n$-punctured Riemann sphere. 
This choice now fixes $P^+_\rho=1$ and we then solve for  $X^-$ using the remaining constraint
\begin{equation}
\partial X^-=\frac{P^i\partial X^i+\Psi^i\partial \Psi^i}{P^+}\, . \label{LC-X-}
\end{equation}
This has the freedom of a constant in the solution for $X^-$ and integrating this out will give conservation of the $+$ component of the external momenta.

With this last gauge fixing and elimination of the remaining constraint, we have reduced to the physical degrees of freedom.  These light-cone gauge variables are the $SO(8)$ bosonic vectors $X^i$ and $P^i$ and the SO(8) fermionic vector $\Psi^{i}$, with the chiral worldsheet action
\begin{eqnarray}
S_{LC} &=& \int d^{2}\rho [
 P^i \bar{\partial}X^i + \frac{1}{2}\Psi^{i}\bar{\partial}\Psi^{i} ]+ S_{J}\, .\label{eqLC}
\end{eqnarray}

\subsection{Interaction point operators, scattering equations and momentum conservation}

One needs to introduce interaction-point operators in light-cone gauge at the $n-2$ points ${\tilde z}_{\alpha}$ at the zeroes of $P^+ (z)$.
These are the light-cone version of picture-changing operators in the covariant RNS amplitude prescription and come from integration over the modes of the worldsheet gravitino $\chi$ and metric $e$ which cannot be gauge-fixed to zero on an $n$-punctured Riemann sphere. In the ordinary light-cone RNS formalism, the interaction-point operator is
\begin{eqnarray}\label{aap2}
U^{LC}_{RNS}&=& (P_z^{i}\Psi_z^{i})|_{\tilde z_\alpha} \left(\frac{\partial P_z^+}{\partial z}\right)^{-\frac{3}{4}}
\end{eqnarray}
where $ (P_z^{i}\Psi_z^{i})|_{\tilde z_\alpha} $ comes from integration over the modes of $\chi$ and the factor of $\left(\frac{\partial P^+}{\partial z}\right)^{-\frac{3}{4}}$ has conformal weight $-\frac{3}{2}$ which cancels the conformal weight of $P^i \Psi^i$.
For the ambitwistor string, one obtains an additional delta function $\delta(P^i P^i)$ from integration over the modes of $e$ together with a factor $\left(\frac{\partial P^+}{\partial z}\right)$ to cancel the conformal weight. So
the ambitwistor light-cone gauge interaction-point operator is
\begin{equation}
U^{LC}_{ambi}:=
\left(\frac{\partial P^+_z}{\partial z}\right)^{\frac14}(P_z^i\Psi^i_z)|_{\tilde z_\alpha}\delta ( P_z^iP_z^i|_{\tilde z_\alpha})
\label{RNS-int}
\end{equation}

Comparing with (\ref{LC-solve}), we see that the delta functions\footnote{Here we use the fact that a fermionic delta function  is $\delta(\eta)=\eta$ for a fermionic variable $\eta$ so that $\delta(P^i_z\Psi^i_z)=P^i_z\Psi^i_z$. } in the interaction-point operator imply the absence of poles in $P^-_z$ and $\Psi^-_z$ at the points where $P^+_z$ vanishes. Furthermore, we can  see that the vanishing of the residues of $P^-$ at the interaction points is equivalent to the scattering equations as follows. First note that on the Riemann sphere, the $P^-$ defined by \eqref{LC-solve} has simple poles at the $z_r$ with residue $k^-_r$.  Although superficially there is a double pole coming from the numerator, the pole in $P^+$ cancels with one of these  to give the residue
\begin{equation}
\mathrm{Res}_{z_r}P^-(z)=\mathrm{Res}_{z_r}\frac{P^iP^i(z)}{P^+(z)}=\frac{k_r^ik_{r}^{i}}{k^+_r}=:k^-_r\, .
\end{equation}
If all the residues at the interaction points vanish, then  we must have
\begin{eqnarray}
 P^{-}(z) = \sum_{r=1}^{N}\frac{k_{r}^{-}}{z-z_{r}}\, .
\end{eqnarray}
However, by definition,  $2P^-(z) P^+(z) = P^j(z) P^j(z)$ so that  $P_mP^m=0$ identically, and hence its residues at $z_r$ vanish which gives the usual form of the scattering equations.  

It is also the case that the sum of all the residues of $P^-$ must vanish as it is a 1-form on the Riemann sphere. Thus the delta functions of residues at the interaction points together imply   $\sum_r k_r^-=0$, which is the final momentum conservation delta function.

Thus the path-integral over the zero-modes of $(X^i,X^-)$ will give the delta function for conservation of the transverse and $+$-components of the momentum, whereas the insertion of $\delta (\mathrm{Res}_{\tilde z_\alpha} P^-)$ at the interaction points will provide the scattering equations and the final momentum conservation delta function.
\subsection{Vertex operators}
To describe Ramond states, we must construct the spin fields. 
 Bosonizing $\Psi^{i}$ in the standard way
\begin{eqnarray}
\frac{1}{\sqrt{2}}(\Psi^{2j} \pm i\Psi^{2j-1}) &=& e^{\pm \tilde{H}_{j}}
\end{eqnarray}
with $\tilde{H}_{i}$ satisfying the OPE
\begin{eqnarray}
\tilde{H}_{i}(z)\tilde{H}_{j}(w) &\rightarrow & \delta_{ij}\log(z-w)
\end{eqnarray}
one can construct the $SO(8)$ chiral and antichiral spin fields of conformal weight $\frac{1}{2}$
\begin{eqnarray}
\tilde{\Sigma}^{a} &=& e^{[\sum_{i=1}^{4}\pm \frac{\tilde{H}_{i}}{2}]} \hspace{2mm},\hspace{2mm} \mbox{for an even number of -'s}\label{ap2}\\
\tilde{\Sigma}^{\dot{a}} &=& e^{[\sum_{i=1}^{4}\pm \frac{\tilde{H}_{i}}{2}]}\hspace{2mm},\hspace{2mm} \mbox{for an odd number of -'s}\label{ap6}
\end{eqnarray}

In terms of the light-cone variables, the spacetime supersymmetry currents are 
\begin{eqnarray}
q_{a} = \frac{i}{\sqrt{2\sqrt{2}}}(\sigma^{i})_{a\dot{b}}\tilde{\Sigma}^{\dot{b}}\frac{P^{i}}{(P^{+})^{\frac{1}{2}}} \hspace{2mm}&,&\hspace{2mm} q_{\dot{a}} = \frac{i}{\sqrt[4]{2}}(P^{+})^{\frac{1}{2}}\tilde{\Sigma}_{\dot{a}}
\end{eqnarray}
which satisfy the OPEs
\begin{eqnarray}
q_{a}(z)q_{b}(w) \rightarrow -\frac{\delta_{ab}P^{-}}{\sqrt{2}(z-w)} \hspace{2mm},\hspace{2mm} q_{a}(z)q_{\dot{b}}(w) \rightarrow -\frac{(\sigma^{i})_{a\dot{b}}P^{i}}{2(z-w)} \hspace{2mm},\hspace{2mm} q_{\dot{a}}(z)q_{\dot{b}}(w) \rightarrow -\frac{\delta_{\dot{a}\dot{b}}P^{+}}{\sqrt{2}(z-w)}\nonumber\\
\end{eqnarray}

The gluon and gluino states are generated at the cylindrical ends of the strings by vertex operators which in light-cone gauge are
\begin{eqnarray}\label{ap1}
V^{LC}_{gluon} &=& \Psi^{i}A_{i}^{I}J_{I}e^{ik^j X^j}  \nonumber\\
V^{LC}_{gluino} &=& (k^{+})^{-\frac{1}{2}}\tilde{\Sigma}^{a}C_a^{I}J_{I} e^{ik^j X^j}
\end{eqnarray}
where 
$A_{i}^{I}$ and $C^{a\,I} = P^{+}B^{a\,I}$ are the gluon and gluino polarizations in the light-cone gauge $A^{+I} = B^{\dot a\,I}=0$.  Here $I$ is a Lie algebra index and $J^I$ a corresponding current algebra.


In terms of these vertex operators and interaction-point operators, the N-point tree-level scattering amplitude prescription is
\begin{eqnarray}
\mathcal{A}^{LC} &=& \langle V_1^{LC}(z_{1})V_2^{LC}(z_{2})U_{ambi}^{LC}(\tilde{z}_{1})V_3^{LC}(z_{3})\ldots U_{ambi}^{LC}(\tilde{z}_{N-2})V_N^{LC}(z_{N}) \rangle
\end{eqnarray}
where $V_r^{LC}$ are the light-cone gauge physical vertex operators defined in \eqref{ap1} which are located at points $z_{r}$ satisfying the scattering equations, and $U^{LC}_{ambi}$ are the interaction-point operators located at points $\tilde{z}_{\alpha}$ satisfying $P^+(\tilde z_\alpha)=0$.

We can see that this is equivalent  to the conventional CHY formula arising from the conventional BRST covariant quantization of the RNS ambitwistor string by comparing this formulation with that given in \citep{Adamo:2013tsa}.  There, the choice of basis of Beltrami differentials is arbitrary and so can be adapted to the interaction points setting $\mu_\alpha= 
\theta_\alpha\bar\delta(P^+_z)$ where $\theta_\alpha=1$  near $\tilde z_\alpha$ and zero near $\tilde z_\beta$ for $\beta\neq \alpha$.  This leads to the formulae given here for the scattering equations at the interaction points.  Furthermore, the insertion points for the picture changing operators are essentially arbitrary, and if inserted at the interaction points they reduce to give $\delta(\beta) (P^i\Psi^i +P^-\Psi^+)$. But in light cone gauge for the external fields,  there will be no $\Psi^-$ for the latter term to contract with, so the $\Psi$ contractions will give the same formulae as for the BRST covariant quantization of the ambitwistors-string. Furthemore, as in the usual RNS string, the path integral over the $(\beta, \gamma)$ ghosts with these insertions will cancel
the path integral over the $(\Psi^+, \Psi^-)$ fields.

\section{Light-cone gauge for twistorial ambitwistor-string}

The heterotic twistorial ambitwistor string, ignoring the current algebra variables in $S_J$, has action
\begin{multline} \label{twistact}
S= \int  \left(w_{\alpha}\bar{\partial}\lambda^{\alpha} + \frac{1}{2}\psi^{m}\bar{\partial}\psi_{m} 
 + h_{\alpha}[(\lambda\gamma^{m}\lambda)(\gamma_{m}w)^{\alpha} -2\lambda^{\alpha}(\lambda w) + \psi^{m}\psi^{n}(\gamma_{m}\gamma_{n}\lambda)^{\alpha} ] +\right.\\
\left.  f(\lambda\gamma^{m}\lambda)\psi_{m} +\tilde e (\frac{1}{2} w_\alpha \partial \lambda^\alpha  - \frac{1}{2} \lambda^\alpha \partial w_\alpha+  \frac{1}{2}\psi^{m}{\partial}\psi_{m}) \right)
\end{multline}
where $h_\alpha$, $f$ and $\tilde e$ are Lagrange multipliers for the constraints $G^\alpha$, $T_F$ and $T_B$ respectively.

In light-cone gauge for the twistorial string, we again completely fix the gauge freedom so that there are no propagating ghosts. Under the SO(8) decomposition of (\ref{dec}), 10d spinors decompose into $SO(8)$ chiral spinors so that
\begin{eqnarray}
\lambda^{\alpha} &=&
\begin{pmatrix}
\lambda^{a} \\
\lambda^{\dot{a}}
\end{pmatrix},
\end{eqnarray}
where $a, \dot{a}$ run from 1 to 8. We will further choose a specific impure 8d spinor $\iota^a$ with $\iota^a\iota_a=1$ and, using the gauge transformations generated by $G^{\alpha}$,  set seven components of $\lambda^{\alpha}$ to zero by requiring 
\begin{equation}
\lambda^a=\lambda^+\iota^a\,.
\end{equation}
With these choices
\begin{align}
P_m&= \lambda\gamma_m \lambda =(-\sqrt{2}\lambda^a\lambda^a,-\sqrt{2}\lambda^{\dot a}\lambda^{\dot a}, 2\lambda^{a}(\sigma^{i})_{a\dot a}\lambda^{\dot a})\nonumber \\&=(-\sqrt{2}(\lambda^+)^2,-\sqrt{2}\lambda^{\dot a}\lambda^{\dot a}, 2\lambda^+(\sigma^{i})_{+\dot a}\lambda^{\dot a})\, ,
\end{align}
where $(\sigma^{i})_{a\dot a}$ are the 8d Pauli matrices and we define $(\sigma^{i})_{+\dot a}=(\sigma^{i})_{a\dot a}\iota^a$.  We can similarly parametrize the external momenta $k^m$ in terms of spinors $\kappa^\alpha=(\kappa^+\iota^a, \kappa^{\dot a})$ with 
\begin{equation}
k^m=(-\sqrt{2}(\kappa^+
)^2,-\sqrt{2}\kappa^{\dot a}\kappa^{\dot a}, 2\kappa^+(\sigma^{i})_{+\dot a}\kappa^{\dot a})\, .
\end{equation}


We now use the transformations in \eqref{eq46} to gauge $\psi^+ =0$ and solve the constraint $T_F=0$ by   expressing $\psi^{-}$ in terms of the transverse components $\psi^{i}$, where $i=1,\ldots,8$, as
\begin{eqnarray}
\psi^{-} = -\sqrt{2}\frac{(\sigma^{i})_{+\dot{a}}\lambda^{\dot{a}}\psi^{i}}{\lambda^{+}}\,.
\end{eqnarray}
We similarly use the constraint $G^\alpha=0$ to solve for the components of $w_{a}$ that are perpendicular to $\iota^a$ as
\begin{equation}
w^a-\iota^a w_+=\sqrt{2}(\delta^{ad}-\iota^a\iota^d)\frac{\lambda^{\dot b}\sigma^i_{+\dot b}\sigma^{i}_{d\dot c}w^{\dot c}}{\lambda^+} +\ldots\, ,
\end{equation}
where $\ldots$ are quadratic terms in $\psi$ that depend on $\lambda$.  This leaves the component $w_+:=w_a\iota^a$ free. We finally use the transformations generated by $T_B$ to 
gauge-fix $\lambda^+=
\frac{i}{\sqrt[4]{2}}$ which fixes the coordinates on the worldsheet.  Since $P^+=-\sqrt{2}(\lambda^+)^2$, this will agree with the standard light-cone gauge choice. Setting $T_B=0$ then allows one to solve for $\partial w_+$ in terms of the other variables.

So in light-cone gauge, the worldsheet action depends only on the bosonic and fermionic transverse worldsheet
variables $(\lambda^{\dot a}, w_{\dot a}, \psi^i)$ of conformal weight $\frac{1}{2}$ and the worldsheet action is
\begin{equation}
S_{LC} = \int d^2 z [ w_{\dot a} \bar\nabla \lambda^{\dot a} + \psi^i \bar\nabla
\psi^i ] +S_J
\end{equation}
where $S_J$ is the current algebra action and $\bar\nabla \equiv \bar\partial - \bar\partial (\log \lambda^+)$. This is defined so that $\bar\nabla \lambda^+ =0$ in any coordinate system.

As before in the RNS case, and in the usual light-cone gauge in string theory, we  identify the momentum $P^+=\frac{\partial\rho}{\partial z}$ where  the Mandelstam map $\rho  (z)$  from the complex plane to the string
worldsheet for $N$-point tree amplitudes is given by \cite{Mandelstam:1973jk,MANDELSTAM197477}
\begin{equation}
\rho(z) = \sum_{r=1}^N k^+_r \log (z-z_r)\, .
\end{equation}
Since $-\sqrt{2}(\lambda^+(z))^2 =P^+(z)= \sum_{r=1}^N \frac{ k^+_r}{ z-z_r}$ in this gauge, we must have
\begin{equation}\label{sqr}
\lambda^+ (z)=\frac{i}{\sqrt[4]{2}}\sqrt {\frac{\partial\rho}{\partial z}} = \frac{i}{\sqrt[4]{2}}\sqrt {\sum_{r=1}^N \frac{k^+_r} {z-z_r}}.
\end{equation}
Thus $\lambda^+(z)$ has square-root cuts at the locations $z=z_r$ and has
square-root zeros at the locations of the $N-2$ interaction-points $z=\tilde z_\alpha$ defined by
\begin{equation}
\left. \frac{\partial\rho}{\partial z}\right|_{z = \tilde z_\alpha}=  \sum_{r=1}^N \frac{P^+_r} {\tilde z_\alpha -z_r}=0.
 \end{equation}

Since the momenta $P^j(z) =2\lambda^+ (\sigma^j)_{+\dot a} \lambda^{\dot a}$ and the supersmmetry generator
$q_{\dot a} = \lambda^+ (\sigma^j)_{+ \dot a} \psi^j$ should not have square-root cuts
anywhere on the worldsheet where $\lambda^+$ is defined by (\ref{sqr}), the transverse worldsheet variables $(\lambda^{\dot a}, \psi^i, w_{\dot a})$ must have square-root cuts at the locations $z=z_r$ and $z=\tilde z_\alpha$.
This is different from the RNS fermionic variable $\Psi^j$ in light-cone gauge which has no square-root cuts at the interaction points $z=\tilde z_\alpha$ and only has square-root cuts at $z=z_r$ for states in the Ramond sector. However, the square-root cuts of $\psi^i$ in this formalism is similar to the Green-Schwarz light-cone fermionic variable which has square-root cuts both at $z=z_r$ and $z=\tilde z_\alpha$. Of course, $\psi^i$ differs from the Green-Schwarz
light-cone fermionic variable in that it is an $SO(8)$ vector instead of an $SO(8)$ spinor, although in our gauge, $(\sigma^i)_{+\dot a}$ can be used to translate.


\subsection{Light-cone gauge vertex operators}
In this subsection we use the light-cone twistor variables to construct physical vertex operators for the gluon and gluino fields $A^{m\,I}$ and $B_\alpha^I$, where $I$ is a Lie algebra index.  We will choose the light-cone  gauge conditions:
$A^{+\,I}  =B_{\dot a}^I = 0$.
The first step to construct the vertex operators is to define the eigenvector of the momentum operator. In this light-cone framework this vertex will have the factor 
\begin{eqnarray}\label{eeq29}
  e^{-w^{\dot a}k_{\dot a}/2\lambda^+} \qquad \mbox{ where } \qquad k_{\dot{a}} := k^{i}(\sigma^{i})_{+\dot{a}}\, .
\end{eqnarray}
  This agrees with $e^{k_ix^i}$ via the incidence relations in light-cone gauge.
In a generic $N$-point correlation function, one will insert $N$ of this type of vertex which will provide the following light-cone equations of motion for the twistor field $\lambda^{\dot a}$
\begin{eqnarray}
\bar{\nabla}\lambda^{\dot{a}} &=&  \frac{1}{2\lambda^{+}}\sum_{r=1}^{N}k^{\dot a}_{r}\delta^{2}(z-z_{r})\label{eq45}
\end{eqnarray}
which implies that 
\begin{eqnarray}\label{eq16}
\lambda^{\dot{a}} &=& \frac{1}{2\lambda^{+}}\sum_{r=1}^{N}\frac{k_{r}^{\dot a}}{z-z_{r}}
\end{eqnarray}
where $\lambda^+$ is defined in (\ref{sqr}).

\vspace{2mm}
Next we will construct the light-cone gauge gluon and gluino vertices using standard bosonization techniques. Defining
\begin{eqnarray}\label{eq17}
\frac{1}{\sqrt{2}}(\psi^{2j} \pm i\psi^{2j-1}) &=& e^{\pm H_{j}} \hspace{2mm},\hspace{2mm} \mbox{for $j=1,2,3,4$}
\end{eqnarray}
where the scalar field $H(z)$ satisfies the OPE
\begin{eqnarray}
H_{i}(z)H_{j}(w) &\rightarrow & \delta_{ij}ln(z-w),
\end{eqnarray}
one can construct the light-cone gauge spin fields as\footnote{Formally, one should also write the so-called cocycles for $\Sigma^{a}$, $\Sigma_{\dot{a}}$. These factors are relevant to get correctly the OPEs between the spin fields and $\psi^{i}$.}
\begin{eqnarray}
\Sigma^{a} &=& e^{\pm\frac{H_{1}}{2}\pm\frac{H_{2}}{2}\pm\frac{H_{3}}{2}\pm\frac{H_{4}}{2}} \hspace{2mm},\hspace{2mm} \mbox{for an even number of -'s}\label{eq18}\\
\Sigma_{\dot{a}} &=& e^{\pm\frac{H_{1}}{2}\pm\frac{H_{2}}{2}\pm\frac{H_{3}}{2}\pm\frac{H_{4}}{2}} \hspace{2mm},\hspace{2mm} \mbox{for an odd number of -'s}\label{eq19}
\end{eqnarray}
which have the usual OPE's, e.g.
\begin{eqnarray}\label{eq100}
\psi^{i}(z) \Sigma_{\dot a}(w) &\rightarrow & \frac{1}{\sqrt{z-w}} (\sigma^i)_{a \dot a} \Sigma^a
\end{eqnarray}
The light-cone gauge vertex operators can now be defined as
\begin{eqnarray}
V_{gluon}(z_{r}) &=& [(\sigma^{j})_{+\dot{b}}\Sigma^{\dot{b}}A^{I}_{j}]J_{I}e^{ - \frac{k_{r}^{i}(\sigma^{i})^{+\dot{a}}w^{\dot{a}}}{2\lambda^{+}}}\label{eeq21}\\
V_{gluino}(z_{r}) &=& \frac{1}{\sqrt{k^+_r}}C^{a\,I}(- \Sigma^{a} + 2 \iota^a \iota^b \Sigma^b) J_{I}e^{ - \frac{k_{r}^{i}(\sigma^{i})^{+\dot{a}}w^{\dot{a}}}{2\lambda^{+}}}\label{eeq22}
\end{eqnarray}
where $A_{j}^{I}$ and $C^{a\,I}$ are the light-cone gauge gluon and gluino polarizations. One can show that the vertex $V = V_{gluon} + V_{gluino}$ is invariant under the light-cone gauge supersymmetries generated by the currents
\begin{eqnarray}
q_{\dot{a}} = \psi^{i}(\sigma^{i})_{+\dot{a}}\lambda^{+} \hspace{2mm}&,&\hspace{2mm} q_{a} = \psi^{i}(\sigma^{i})_{a\dot{a}}\lambda^{\dot{a}} - 2\psi^{i}(\sigma^{i})_{+\dot{a}}\lambda^{\dot{a}}\delta_{a+}
\end{eqnarray}
which satisfy the OPEs
\begin{eqnarray}
q_{a}(z)q_{b}(w) \rightarrow \frac{\delta_{ab}\lambda^{\dot{c}}\lambda^{\dot{c}}}{z-w} \hspace{2mm},\hspace{2mm} q_{a}(z)q_{\dot{a}}(w)\rightarrow -\frac{\lambda^{+}(\sigma^{i})_{+\dot{a}}\lambda^{\dot{a}}}{z-w}\hspace{2mm},\hspace{2mm} q_{\dot{a}}(z)q_{\dot{b}}(w) \rightarrow \frac{\delta_{\dot{a}\dot{b}}(\lambda^{+})^{2}}{z-w}
\end{eqnarray}
\subsection{Light-cone gauge scattering amplitudes}

To compute scattering amplitudes, one first needs to introduce interaction-point operators located at the zeros $\tilde z_\alpha$ of $\frac{\partial \rho}{\partial  z}$. The light-cone gauge scattering amplitudes can then be computed using the prescription
\begin{eqnarray}\label{eeq20}
\mathcal{A} &=& \langle V_1(z_{1})V_2(z_{2})U_{int}(\tilde{z}_{{1}})V_3(z_{3})\ldots U_{int}(\tilde{z}_{{N-2}})V_N(z_{N}) \rangle
\end{eqnarray}
where $V_r$, $U_{int}$ are the physical vertices and interaction-point operators.

Since $\psi^j (z)$ should have square-root cuts at $z=\tilde z_\alpha$, the interaction-point operators should contain the spin field $\Sigma_{\dot\alpha}$ and will be defined as 
\begin{eqnarray}\label{eq15}
U_{int}(\tilde{z}_{\alpha}) &=& (\tilde\lambda^{\dot{a}}\Sigma^{\dot{a}}) \delta (\tilde\lambda^{\dot b}\tilde\lambda^{\dot b})(\frac{\partial^2 \rho}{\partial z^2})^{\frac{1}{4}}
\end{eqnarray}
where 
$$\tilde\lambda^{\dot a}\equiv \frac{i}{\sqrt[4]{2}}\sqrt{ \frac{\partial\rho}{\partial z}}\lambda^{\dot a} =
\frac{1}{2}(\sigma^{i})^{\dot{a}+}\sum_{r=1}^{N}\frac{k_{r}^{i}}{\tilde{z}_{\alpha}-z_{r}}$$
and, as in (\ref{RNS-int}), the factor of $(\frac{\partial^2 \rho}{\partial z^2})^{\frac{1}{4}}$ carries conformal weight of $\frac{1}{2}$ which cancels the conformal weight of $ (\tilde\lambda^{\dot{a}}\Sigma^{\dot{a}}) \delta (\tilde\lambda^{\dot b}\tilde\lambda^{\dot b})$. In principle, it should be possible to derive this interaction-point operator from gauge-fixing the covariant action of (\ref{twistact}), but we do not yet see how to derive (\ref{eq15}) in this manner.


We will now show the equivalence of the N-point correlation function given by \eqref{eeq20} and the one obtained in the standard RNS ambitwistor string by finding an identification of the variables in the two models. 

\vspace{2mm}
We start by relating the light-cone gauge RNS fermionic vector $\Psi^{i}$ with the fermions in
the twistorial description by the identification
\begin{eqnarray}\label{eq78}
\Psi^{i} &=& (\sigma^{i})_{+\dot{a}}\Sigma^{\dot{a}}
\end{eqnarray}
where $\Sigma^{\dot{a}}$ is the spin field constructed out of the light-cone gauge supertwistor fermionic vector $\psi^{i}$ as explained in \eqref{eq19}. Then the spin field obtained from $\Psi^{i}$ can be identified to $\psi^{i}$ through the relation
\begin{eqnarray}\label{eq79}
\tilde{\Sigma}_{\dot{a}} &=& (\sigma^{i})_{+\dot{a}}\psi^{i}
\end{eqnarray}
where eqn. \eqref{eq79} is a direct consequence of the definition of $\tilde{\Sigma}_{\dot{a}}$ in \eqref{ap6} and the bosonization of $\psi^{i}$ in \eqref{eq17}.

\vspace{2mm}
Eqn. \eqref{eq78} can be used to relate the two gluon vertex operators corresponding to both models as follows
\begin{eqnarray}
V^{RNS}_{gluon} = \Psi^{i}A^{I}_{i}J_{I} \hspace{2mm}&,&\hspace{2mm} V^{Twistor}_{gluon} = (\sigma^{i})_{+\dot{a}}A^{I}_{i}J_{I}\Sigma^{\dot{a}}
\end{eqnarray}
Using the twistor identity $P^{m} = \lambda\gamma^{m}\lambda$, one can immediately relate the interaction-point operators
\begin{eqnarray}
 U_{int}^{RNS} =P_{i}\Psi^{i} \delta(P_iP^i)\left(\frac{\partial P^+}{\partial z}\right)^{\frac{1}{4}}
 \hspace{2mm}&,&\hspace{2mm}U^{Twistor}_{int} =
  (\tilde\lambda^{\dot{a}}\Sigma_{\dot{a}}) \delta (\tilde\lambda_{\dot\beta}\tilde\lambda^{\dot\beta})(\frac{\partial^2 \rho}{\partial z^2})^{\frac{1}{4}}
\end{eqnarray}
Furthermore, eqn. \eqref{eq79} allows us easily to relate the supersymmetry generators associated to both models
\begin{eqnarray}
q^{RNS}_{\dot{a}} = \frac{i}{\sqrt[4]{2}}(P^{+})^{\frac{1}{2}}\tilde{\Sigma}_{\dot{a}} \hspace{2mm}&,&\hspace{2mm} q^{Twistor}_{\dot{a}} = \lambda^{+}(\sigma^{i})_{+\dot{a}}\psi^{i}
\end{eqnarray}
And since the gluino vertex is obtained from $V_{gluon}$ by supersymmetry in both models, one has an analogous relation between the fermionic vertices.

Thus the $N$-point tree amplitude prescription of (\ref{eeq20}) in this twistorial ambitwistor formalism is equivalent to the $N$-point tree amplitude prescription in the light-cone ambitwistor version of the RNS formalism.

\section{Discussion}
We have  seen that the 10d twistorial ambitwistor-string can be quantized in light cone gauge so as to generate formulae for amplitudes.  These formulae are most simply compared with the CHY formulae \cite{Cachazo:2013hca} via the RNS model for ambitwistor-strings \cite{Mason:2013sva} quantized in light-cone gauge.  In the RNS light-cone gauge, we have seen that the interaction point operators play the role of imposing the scattering equations and the picture-changing operators.  These eliminate the spurious singularities in the worldsheet fields that have been obtained by solving the constraints.  

We find that the fermionic vector $\psi_m$ of the 10d twistor model is not  naturally identified with the $\Psi_m$ of the RNS model.  Instead, in their light-cone gauge reduction they live in each-other's Ramond sector.  This is something that can be inferred covariantly from the form of the supersymmetry generator which is $(\psi\lambda)_\alpha$ in the twistor model whereas it is constructed from the Ramond sector in the RNS model. This presents a challenge for the construction of covariant vertex operators.  

The light-cone gauge for the 10d twistor model introduces square-roots into the momentum spinor $\lambda^\alpha$ which is constructed rather directly and non-covariantly from the ambitwistor momentum $P_m$.  Covariant quantization of the twistor-string \cite{ Witten:2003nn, Berkovits:2004hg,Skinner:2013xp} and twistorial ambitwistor-string models \cite{Geyer:2014fka,Geyer:2018xgb,Geyer:2019ayz} in respectively 4, 6 and 10/11 dimensions leads to rational expressions for the spinor constituents of $P_m$.  The latter formulae are based on the \emph{polarized} scattering equations which incorporate polarization data into the constituent spinors. It is to be hoped that a covariant quantization can be found for the 10d twistor model studied here that manifests some of these features with a rational $\lambda^\alpha$.

\section{Acknowledgments}
NB acknowledges FAPESP grants 2016/01343-7 and 2014/18634-9 and CNPq grant 300256/94-9 for partial financial support. MG acknowledges FAPESP grants 15/23732-2 and 18/10159-0 for financial support. LJM acknowledge support from the EPSRC grant EP/M018911/1. MG would also like to thank the hospitality of Mathematical Institute at Oxford where much of this research was carried out and Perimeter Institute for their hospitality during the preparation of the manuscript. This project has received support from the European Union\textquotesingle s Horizon 2020 research and innovation programme under the Marie Sk\l odowska-Curie grant agreement No. 764850, SAGEX.


\begin{thebibliography}{10}

\bibitem{Mason:2013sva}
L.~Mason and D.~Skinner, ``{Ambitwistor strings and the scattering
  equations},'' \href{http://dx.doi.org/10.1007/JHEP07(2014)048}{{\em JHEP}
  {\bf 07} (2014)  048},
\href{http://arxiv.org/abs/1311.2564}{{\tt arXiv:1311.2564 [hep-th]}}.

\bibitem{Berkovits:2013xba}
N.~Berkovits, ``{Infinite Tension Limit of the Pure Spinor Superstring},''
  \href{http://dx.doi.org/10.1007/JHEP03(2014)017}{{\em JHEP} {\bf 03} (2014)
  017},
\href{http://arxiv.org/abs/1311.4156}{{\tt arXiv:1311.4156 [hep-th]}}.

\bibitem{Casali:2015vta}
E.~Casali, Y.~Geyer, L.~Mason, R.~Monteiro, and K.~A. Roehrig, ``{New
  Ambitwistor String Theories},''
  \href{http://dx.doi.org/10.1007/JHEP11(2015)038}{{\em JHEP} {\bf 11} (2015)
  038},
\href{http://arxiv.org/abs/1506.08771}{{\tt arXiv:1506.08771 [hep-th]}}.

\bibitem{Cachazo:2013hca}
F.~Cachazo, S.~He, and E.~Y. Yuan, ``{Scattering of Massless Particles in
  Arbitrary Dimensions},''
  \href{http://dx.doi.org/10.1103/PhysRevLett.113.171601}{{\em Phys. Rev.
  Lett.} {\bf 113} (2014) no.~17, 171601},
\href{http://arxiv.org/abs/1307.2199}{{\tt arXiv:1307.2199 [hep-th]}}.

\bibitem{Cachazo:2013iea}
F.~Cachazo, S.~He, and E.~Y. Yuan, ``{Scattering of Massless Particles:
  Scalars, Gluons and Gravitons},''
  \href{http://dx.doi.org/10.1007/JHEP07(2014)033}{{\em JHEP} {\bf 07} (2014)
  033},
\href{http://arxiv.org/abs/1309.0885}{{\tt arXiv:1309.0885 [hep-th]}}.

\bibitem{Cachazo:2014xea}
F.~Cachazo, S.~He, and E.~Y. Yuan, ``{Scattering Equations and Matrices: From
  Einstein To Yang-Mills, DBI and NLSM},''
  \href{http://dx.doi.org/10.1007/JHEP07(2015)149}{{\em JHEP} {\bf 07} (2015)
  149},
\href{http://arxiv.org/abs/1412.3479}{{\tt arXiv:1412.3479 [hep-th]}}.

\bibitem{Witten:2003nn}
E.~Witten, ``{Perturbative gauge theory as a string theory in twistor space},''
  \href{http://dx.doi.org/10.1007/s00220-004-1187-3}{{\em Commun.Math.Phys.}
  {\bf 252} (2004)  189--258},
\href{http://arxiv.org/abs/hep-th/0312171}{{\tt arXiv:hep-th/0312171
  [hep-th]}}.

\bibitem{Berkovits:2004hg}
N.~Berkovits, ``{An Alternative string theory in twistor space for N=4
  superYang-Mills},''
  \href{http://dx.doi.org/10.1103/PhysRevLett.93.011601}{{\em Phys.Rev.Lett.}
  {\bf 93} (2004)  011601},
\href{http://arxiv.org/abs/hep-th/0402045}{{\tt arXiv:hep-th/0402045
  [hep-th]}}.

\bibitem{Skinner:2013xp}
D.~Skinner, ``{Twistor Strings for N=8 Supergravity},''
\href{http://arxiv.org/abs/1301.0868}{{\tt arXiv:1301.0868 [hep-th]}}.

\bibitem{Adamo:2013tsa}
T.~Adamo, E.~Casali, and D.~Skinner, ``{Ambitwistor strings and the scattering
  equations at one loop},''
  \href{http://dx.doi.org/10.1007/JHEP04(2014)104}{{\em JHEP} {\bf 04} (2014)
  104},
\href{http://arxiv.org/abs/1312.3828}{{\tt arXiv:1312.3828 [hep-th]}}.

\bibitem{Azevedo:2017lkz}
T.~Azevedo and O.~T. Engelund, ``{Ambitwistor formulations of R$^{2}$ gravity
  and (DF)$^{2}$ gauge theories},''
  \href{http://dx.doi.org/10.1007/JHEP11(2017)052}{{\em JHEP} {\bf 11} (2017)
  052},
\href{http://arxiv.org/abs/1707.02192}{{\tt arXiv:1707.02192 [hep-th]}}.

\bibitem{BERKOVITS199045}
N.~Berkovits, ``A supertwistor description of the massless superparticle in
  ten-dimensional superspace,''
  \href{http://dx.doi.org/https://doi.org/10.1016/0370-2693(90)91047-F}{{\em
  Physics Letters B} {\bf 247} (1990) no.~1, 45 -- 49}.
  \url{http://www.sciencedirect.com/science/article/pii/037026939091047F}.

\bibitem{Carabine:2018kdg}
N.~Carabine and R.~A. Reid-Edwards, ``{An Alternative Perspective on
  Ambitwistor String Theory},''
\href{http://arxiv.org/abs/1809.05177}{{\tt arXiv:1809.05177 [hep-th]}}.

\bibitem{Witten:1985nt}
E.~Witten, ``{Twistor - Like Transform in Ten-Dimensions},''
\href{http://dx.doi.org/10.1016/0550-3213(86)90090-8}{{\em Nucl.Phys.} {\bf
  B266} (1986)  245}.

\bibitem{Mandelstam:1985ww}
S.~Mandelstam, ``{THE INTERACTING STRING PICTURE AND FUNCTIONAL INTEGRATION},''
  in {\em {Workshop on Unified String Theories Santa Barbara, California, July
  29-August 16, 1985}}, pp.~46--102.
\newblock
1985.
\newblock

\bibitem{MANDELSTAM1973205}
S.~Mandelstam, ``Interacting-string picture of dual-resonance models,''
  \href{http://dx.doi.org/https://doi.org/10.1016/0550-3213(73)90622-6}{{\em
  Nuclear Physics B} {\bf 64} (1973)  205 -- 235}.
  \url{http://www.sciencedirect.com/science/article/pii/0550321373906226}.

\bibitem{Mandelstam:1973jk}
S.~Mandelstam, ``{Interacting String Picture of Dual Resonance Models},''
\href{http://dx.doi.org/10.1016/0550-3213(73)90622-6}{{\em Nucl. Phys.} {\bf
  B64} (1973)  205--235}.

\bibitem{MANDELSTAM197477}
S.~Mandelstam, ``Interacting-string picture of the neveu-schwarz-ramond
  model,''
  \href{http://dx.doi.org/https://doi.org/10.1016/0550-3213(74)90127-8}{{\em
  Nuclear Physics B} {\bf 69} (1974) no.~1, 77 -- 106}.
  \url{http://www.sciencedirect.com/science/article/pii/0550321374901278}.

\bibitem{Geyer:2014fka}
Y.~Geyer, A.~E. Lipstein, and L.~J. Mason, ``{Ambitwistor Strings in Four
  Dimensions},'' \href{http://dx.doi.org/10.1103/PhysRevLett.113.081602}{{\em
  Phys. Rev. Lett.} {\bf 113} (2014) no.~8, 081602},
\href{http://arxiv.org/abs/1404.6219}{{\tt arXiv:1404.6219 [hep-th]}}.

\bibitem{Geyer:2018xgb}
Y.~Geyer and L.~Mason, ``{Polarized Scattering Equations for 6D
  Superamplitudes},''
  \href{http://dx.doi.org/10.1103/PhysRevLett.122.101601}{{\em Phys. Rev.
  Lett.} {\bf 122} (2019) no.~10, 101601},
\href{http://arxiv.org/abs/1812.05548}{{\tt arXiv:1812.05548 [hep-th]}}.

\bibitem{Geyer:2019ayz}
Y.~Geyer and L.~Mason, ``{The M-theory S-matrix},''
\href{http://arxiv.org/abs/1901.00134}{{\tt arXiv:1901.00134 [hep-th]}}.

\end{thebibliography}

\providecommand{\href}[2]{#2}\begingroup\raggedright\endgroup

\end{document}